\documentclass[aps,prb,preprint,superscriptaddress]{revtex4-2}
\usepackage{amsmath,amssymb}
\usepackage{graphicx}  % needed for figures
\usepackage{dcolumn}   % needed for some tables
\usepackage{bm}        % for math
 \usepackage{float}
\usepackage{amsfonts}
\usepackage{gensymb}
\usepackage{upgreek}
\usepackage{mathrsfs} 
 \usepackage{hyperref}

\hypersetup{
    colorlinks=true,
    linkcolor=blue,
    filecolor=magenta,      
    urlcolor=blue,
}
 
\usepackage{blindtext,tikz}
\usetikzlibrary{quantikz}
\usetikzlibrary{calc}

\begin{document}

% Full title of the paper (Capitalized)
\title{The Fractional Orbital Chern Hall Effect}

\author{Christopher Sims}
\affiliation{Elmore Family school of Electrical and Computer Engineering, Purdue University}
\email{Sims58@purdue.edu}
\date{\today}
\begin{abstract}
The fractional quantum hall effect (FQHE) is a milestone of modern day physics, its disovery paved the way for the study of fractional charges which do not obey abelian physics. However, all FQHE require an external magnetic field in order for there to be fractionally charged electrons. This work shows a theoretical study of strongly interaction electrons in the Kagome lattice with magnetism. Under optimal conditions, a band gap opens on the surface of the material which hosts fractional Fermions. These fractional Fermions form composite quasi-electrons without the need for external magnetic field. These states are predicted to host the FQHE without the need for an external magnetic field.
\end{abstract}

\maketitle

%%%%%%%%%%%%%%%%%%%%%%%%%%%%%%%%%%%%%%%%%%
\section{Introduction}

The quantum hall effect was discovered in 1980, this discovery was provided an avenue to study the effect of magnetism on electrons in a system\cite{Wakabayashi1978,Klitzing1980}. Shortly after the quantum hall effect was observed, the fractional quantum hall effect was unveiled\cite{Tsui1982}. This system differs from the QHE because electrons are now confined to 2D and the electrons form Laughlin states\cite{Laughlin1983} under applied magnetic fields\cite{Jain1989,Halperin1993}. While these two discoveries are essential to the understanding of electrons in a magnetic field\cite{Stormer1999}, the study of electrons without a magnetic field was not discovered until 2007\cite{Koenig2007} (predicted in 2004\cite{Bernevig2006,Bernevig2006a}). With this discovery, it is found that intrinsic magnetism from spin interactions in a lattice can lead to a quantum hall effect, commonly referred to as the quantum spin hall effect (QSHE)\cite{Koenig2008}. An extension of this effect is the quantum anomalous hall effect with originates from the berry curvature (Chern insulators), in which there is quantized edge mode without magnetic field\cite{Chang2013,Liu2016}.

With all of these discoveries there still exists a ``missing'' piece to the puzzle. A fractional hall effect without an applied magnetic field has not been experimentally discovered to date. The main issue is there is no experimental evidence that fractional Fermions can be formed without the need for an applied magnetic field\cite{Tang2011}. It has been predicted that from a strongly interaction Haldane model with spin orbit coupling\cite{Rod2015} that a gapped Chern insulator can form. Inside the gap, these newly predicted states host fractional Dirac fermions which do not need a magnetic field in order to exist\cite{Neupert2011,Neupert2011a}. The origin of these fractional Dirac states come from the fractionally charged hybridized orbitals which have a fractional Chern number associated with them\cite{Neupert2011a,Neupert2012,Neupert2012a}.

It has been predicted that fractional Chern insulators are states beyond the magnetically supported Laughlin states which make up the FQHE\cite{Grushin2012}. Therefore, throughout this work, fractional Chern Fermions will refer to fractional non-Laughlin states such as the fractional Dirac Fermion. Fractional Chern Fermions have been observed in van-der-walls heterostructures, however, they need a magnetic field in order to exhibit the fractional hall effect which still exists under the genus of the FQHE\cite{Xie2021,Cheng2019}. Recently, it has been predicted that strongly gapped Kagome lattices can host fractional Dirac fermions without the need for an external magnetic field\cite{Yin2020,Sims2022}.

This work predicts the fractional orbital Chern hall effect, a hall effect with fractional filling which can be observed without the need for an external applied magnetic field. This prediction is support by Hamiltonian simulation and theoretical modeling. In addition, materials and experimental methods are put forward to observe this effect. 
%%%%%%%%%%%%%%%%%%%%%%%%%%%%%%%%%%%%%%%%%%
\section{Materials and Methods}
In order to model the edge states of the Chern gapped insulator, a spin full tight binding Haldane\cite{Kane2005,Liu2011} model for the Kagome lattice\cite{Ezawa2013,Ezawa2013a,Kim2016} is constructed:
\begin{equation}
\begin{aligned}
H = & {} -t \sum_{\langle i,j \rangle \alpha } e^{2\pi i \Phi_{ij}} c^\dagger_{i\alpha}c_{j\alpha}\\ 
&-\mu \sum_{ i \alpha } c^\dagger_{i\alpha}c_{j\alpha} \\
&+i\frac{\lambda_{SO}}{3\sqrt{3}} \sum_{ \ll i,j \gg \alpha } s^\alpha_z e^{2\pi i \Phi_{ij}} v_{ij} c^\dagger_{i\alpha}c_{j\alpha} \\
&+\lambda_V \sum_{ i \alpha } t^{i}_{z} c^\dagger_{i\alpha}c_{j\alpha} \\
&+\lambda_X \sum_{ i \alpha }  s^\alpha_z c^\dagger_{i\alpha}c_{j\alpha} \\
&+\lambda_{SX} \sum_{ i \alpha } t^{i}_{z} s^\alpha_z c^\dagger_{i\alpha}c_{j\alpha} \\
&+\lambda_{SSO} \sum_{ \ll i,j \gg \alpha} t^{i}_{z} s^\alpha_z e^{2\pi i \Phi_{ij}} v_{ij} c^\dagger_{i\alpha}c_{j\alpha}
\end{aligned}
\end{equation}

The notations are as follows: $s_z =\uparrow  \downarrow $, $t_z = A,B$, $ s^\alpha_z = \pm 1$.
Where the terms $t$ is the transfer energy. The second term $\mu$ is the chemical potential. The third term $\lambda_{SO}$ is the spin orbit coupling. The fourth therm $\lambda_V$ is the staggered sublattice which is effected by external electric field. The fifth term $\lambda_X$ is the exchange term. With the sixth and seventh terms being the staggered exchange $\lambda_SX$ and the staggered spin terms $\lambda_SSO$.

%%%%%%%%%%%%%%%%%%%%%%%%%%%%%%%%%%%%%%%%%%
\section{Results \& Discussion}

\begin{figure}[H]
\begin{centering}
\includegraphics[width=10.5 cm]{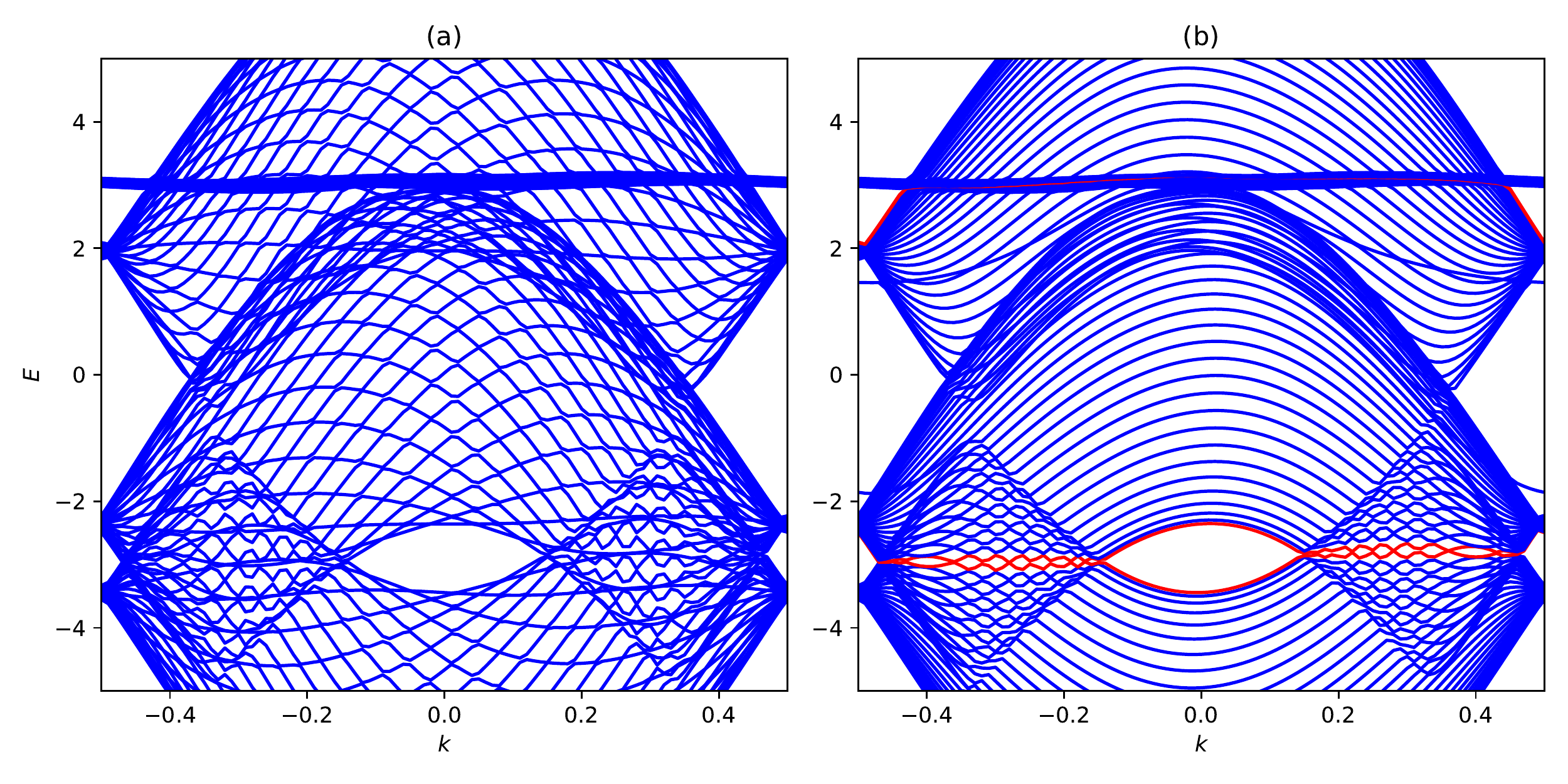}
\caption{Surface edge spectrum of the Kagome Hamiltonian with $t=2.7$,$\mu = 0.2$,$\lambda_{SO}=0.1t$. Blue indicates bulk states and red indicates surface states. (\textbf{a}) Bulk (\textbf{b}) surface\label{SS}}
\end{centering}
\end{figure}

A Kagome Lattice is constructed in the tight binding regime in order to confirm previous DFT results which predict gapped Chern Fermions in the lattice with intrinsic magnetism. In the full tight binding model there exists terms which include correlation and spin correlation effects, these terms ($\lambda_{SX}$,$\lambda_{SSO}$,$\lambda_{X}$,$\lambda_{V}$) are set to zero. The model parameters are set to $t = 2.7$, $\lambda_{SO} = 0.1t$, and $\mu = 0.2$. From this tight binding model the bulk band structure for the Kagome is constructed in the $\Gamma$-$K$-$M$-$K$-$\Gamma$ direction. The band structure reveals a gapped state at the $K$ point with a secondary band gap along the $K$-$\Gamma$ line (Fig. \ref{A1}). In order to see the surface states, a stacked model is constructed to model the bulk and surface states. In the bulk [Fig. \ref{SS}(A)], there exists no band gap in the Kagome lattice. However, on the surface  [Fig \ref{SS}(B)], a gap forms with surface states. there exists multiple crossings due to the strongly interacting nature of the Kagome bands. A Wannier charge center calculation shows high Chern numbers (n>1) (Fig \ref{A2}) which is critical for the formation of fractionally charged Fermions\cite{Regnault2011}. The Wannier charge center analysis does not show fractional charge since these states are the bulks states projected onto the edge, as opposed to the gapped states that form on the surface.

\subsection{Hall Resistance}
In order to model the fractional hall effect in the material candidates, composite fermion theory is utilized in order to construct the fractional filling factors with their associated Chern bands. In this work the parameterization of the field $p$ is characterized by the intrinsic magnetic field in the material. Fractional Fermion theory shows that this parameter scales with the Chern number due to stronger interactions with the intrinsic fields. In high Chern number materials such as Kagome magnets, the base Chern number bands in the ungapped bulk can reach Chern numbers of $C = \geq 20$. Here For the landau levels, six are selected for the base bands with fractionalization occurring on the surface. 

\begin{figure}[!ht]
\begin{centering}
\includegraphics[width=10.5 cm]{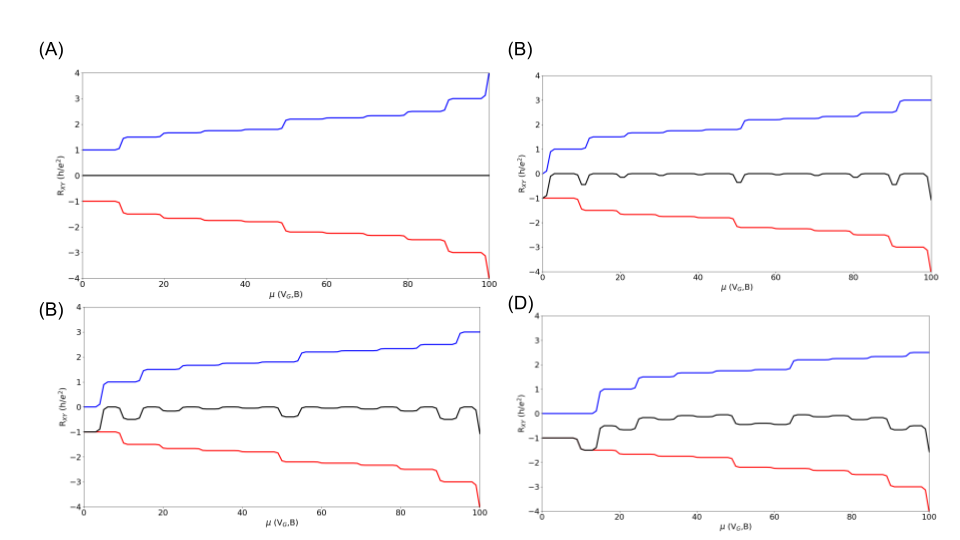}
\caption{$R_{XY}$ as a function of $\mu$ the chemical potential, blue indicates $C_\uparrow$, red indicates $C_\downarrow$, \label{hall}}
\end{centering}
\end{figure}   

In the case of Kagome magnets RMn$_6$Ge$_6$ (R = ) the states inside the gap form fractionally charged Dirac Fermions on the surface which are spin polarized. Although the fractional states are formed from orbital Chern effects the surface (FOCHE), these state are spin polarized with opposite chirality, this particular hall effect is most accurately described by the term Fractional Quantum spin Hall effect (FQSHE). Due to the fact that both the positive ($C_+$) and negative ($C_-$) Chern bands exist at the same energy level there needs to be a penalty function in the chemical potential of each band added in order to observe the fractional filling in the hall effect, otherwise there will be no experimental signature [Fig. \ref{hall}(A)]. In the spinfull kagome lattice, the spin Chern numbers $C_{spin} = \frac{1}{2}(C_\uparrow - C_\downarrow$ are slightly offset due to spliting caused by the intrinsic magnetic field. By apply asymmetric gates to the top ($V_{top}$) and the bottom($V_{bottom}$) of a mesoscopic sample. $\mu(V_g,B)$ shows a parameterization of the spin polarized $C_\uparrow$ and $C_\downarrow$ hall conductivities with respect to the net gate voltage on each state from asymmetric gating. $\mu_{offset}$ is increased only for $C_\uparrow$ with of 5,10, and 15 showing an increase in hall plateaus with gating [Fig. \ref{hall}(C-D)].

%%%%%%%%%%%%%%%%%%%%%%%%%%%%%%%%%%%%%%%%%%
\section{Conclusions}

In conclusion, it is predicted that Chern gapped insulators exhibit the FQHE without the need for external magnetic field. This new effect is termed the Fractional Quantum Spin Hall Effect (FQSHE) for gapped, spin polarized Dirac Fermions. This effect is predicted to exist in Kagome materials RMn$_6$Ge$_6$ under dual gate conditions. This work provides motivation for the study of fractional topology without the need for an external magnetic field in kagome magnets. In future work, the Kagome lattice model can be extended to include more exchange parameters and be extended to see the Hofstadter’s response in the presence of a magnetic field.

%%%%%%%%%%%%%%%%%%%%%%%%%%%%%%%%%%%%%%%%%%
%% Optional

\appendix*
The bulk band structure shows good agreement with previous experimental and theoretical investigations into Kagome materials. The flat band is missing due to the fact that only 3 orbitals are used for this calculation.
\begin{figure}[H]
\begin{centering}
\includegraphics[width=10.5 cm]{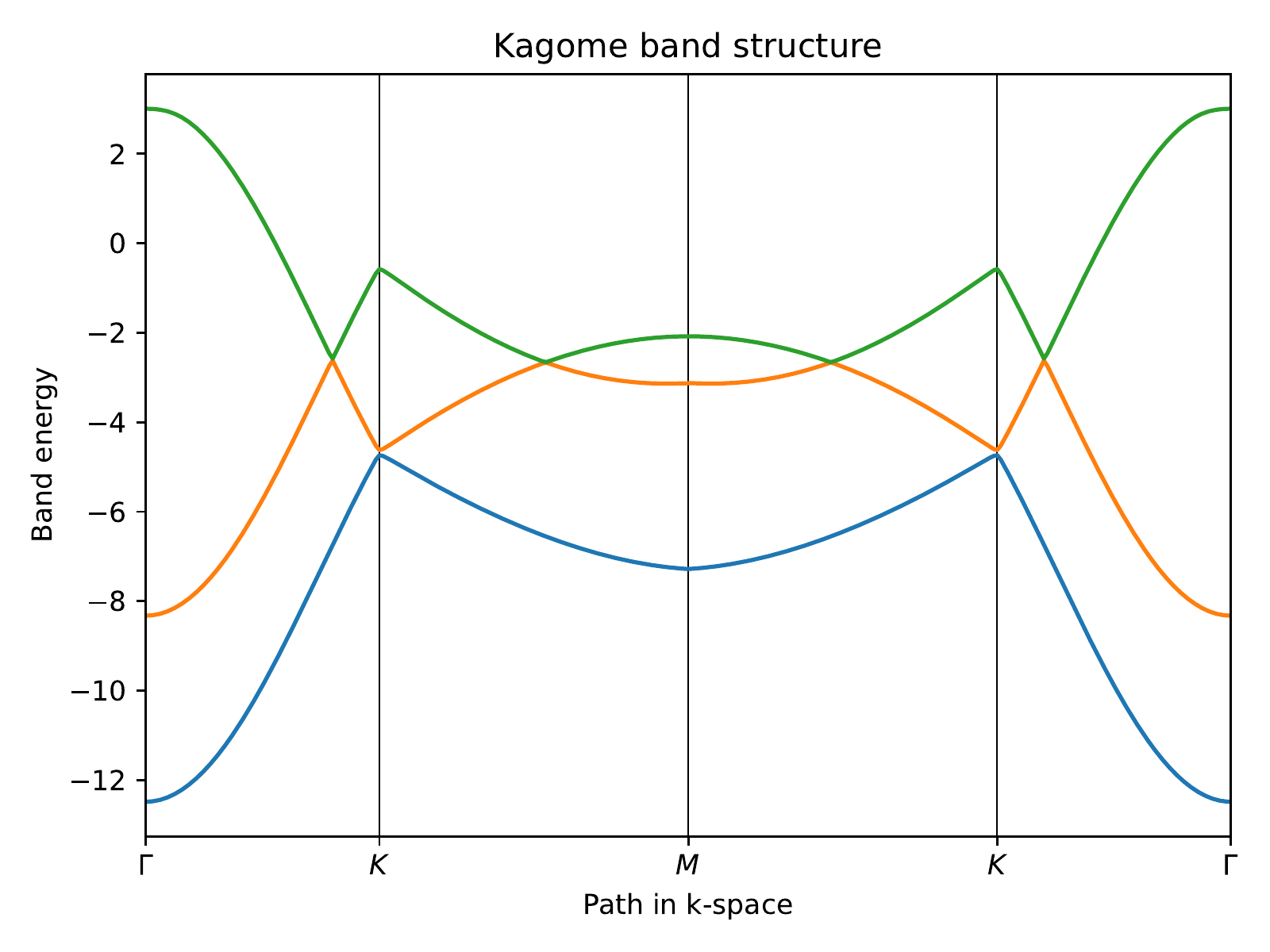}
\caption{Bulk band dispersion of the Kagome lattice with 3 bands with $t = 2.7$ and $\lambda_{SO} = 0.1t$\label{A1}}
\end{centering}
\end{figure}   
\unskip

The Wilson loop analysis shows good agreement with previous studies which found high Chern numbers in the Bulk tight binding Hamiltonians. Fractional Chern numbers are not seen due to the fact that there is no surface component in this edge model
\begin{figure}[H]
\begin{centering}
\includegraphics[width=10.5 cm]{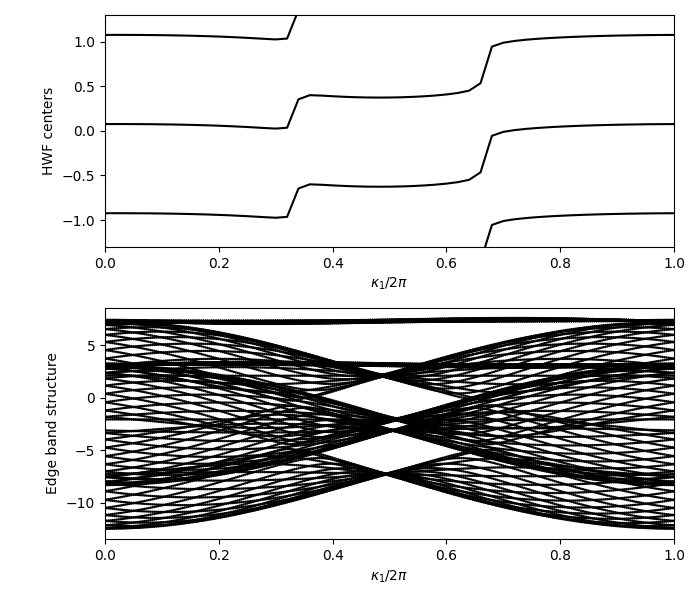}
\caption{(A) Wilson loop calculation and (B) edge bands of the bulk states of the tight binding model\label{A2}}
\end{centering}
\end{figure}

%%%%%%%%%%%%%%%%%%%%%%%%%%%%%%%%%%%%%%%%%
%=====================================
% References, variant A: external bibliography
%=====================================
%\bibliography{ref_FOCHE}

%aipnum4-2.bst 2019-01-14 (MD) hand-edited version of apsrev4-1.bst
%Control: key (0)
%Control: author (8) initials jnrlst
%Control: editor formatted (1) identically to author
%Control: production of article title (-1) disabled
%Control: page (0) single
%Control: year (1) truncated
%Control: production of eprint (0) enabled
\begin{thebibliography}{30}%
\makeatletter
\providecommand \@ifxundefined [1]{%
 \@ifx{#1\undefined}
}%
\providecommand \@ifnum [1]{%
 \ifnum #1\expandafter \@firstoftwo
 \else \expandafter \@secondoftwo
 \fi
}%
\providecommand \@ifx [1]{%
 \ifx #1\expandafter \@firstoftwo
 \else \expandafter \@secondoftwo
 \fi
}%
\providecommand \natexlab [1]{#1}%
\providecommand \enquote  [1]{``#1''}%
\providecommand \bibnamefont  [1]{#1}%
\providecommand \bibfnamefont [1]{#1}%
\providecommand \citenamefont [1]{#1}%
\providecommand \href@noop [0]{\@secondoftwo}%
\providecommand \href [0]{\begingroup \@sanitize@url \@href}%
\providecommand \@href[1]{\@@startlink{#1}\@@href}%
\providecommand \@@href[1]{\endgroup#1\@@endlink}%
\providecommand \@sanitize@url [0]{\catcode `\\12\catcode `\$12\catcode
  `\&12\catcode `\#12\catcode `\^12\catcode `\_12\catcode `\%12\relax}%
\providecommand \@@startlink[1]{}%
\providecommand \@@endlink[0]{}%
\providecommand \url  [0]{\begingroup\@sanitize@url \@url }%
\providecommand \@url [1]{\endgroup\@href {#1}{\urlprefix }}%
\providecommand \urlprefix  [0]{URL }%
\providecommand \Eprint [0]{\href }%
\providecommand \doibase [0]{https://doi.org/}%
\providecommand \selectlanguage [0]{\@gobble}%
\providecommand \bibinfo  [0]{\@secondoftwo}%
\providecommand \bibfield  [0]{\@secondoftwo}%
\providecommand \translation [1]{[#1]}%
\providecommand \BibitemOpen [0]{}%
\providecommand \bibitemStop [0]{}%
\providecommand \bibitemNoStop [0]{.\EOS\space}%
\providecommand \EOS [0]{\spacefactor3000\relax}%
\providecommand \BibitemShut  [1]{\csname bibitem#1\endcsname}%
\let\auto@bib@innerbib\@empty
%</preamble>
\bibitem [{\citenamefont {ichi Wakabayashi}\ and\ \citenamefont
  {Kawaji}(1978)}]{Wakabayashi1978}%
  \BibitemOpen
  \bibfield  {author} {\bibinfo {author} {\bibfnamefont {J.}~\bibnamefont {ichi
  Wakabayashi}}\ and\ \bibinfo {author} {\bibfnamefont {S.}~\bibnamefont
  {Kawaji}},\ }\href {https://doi.org/10.1143/jpsj.44.1839} {\bibfield
  {journal} {\bibinfo  {journal} {Journal of the Physical Society of Japan}\
  }\textbf {\bibinfo {volume} {44}},\ \bibinfo {pages} {1839} (\bibinfo {year}
  {1978})}\BibitemShut {NoStop}%
\bibitem [{\citenamefont {v.~Klitzing}, \citenamefont {Dorda},\ and\
  \citenamefont {Pepper}(1980)}]{Klitzing1980}%
  \BibitemOpen
  \bibfield  {author} {\bibinfo {author} {\bibfnamefont {K.}~\bibnamefont
  {v.~Klitzing}}, \bibinfo {author} {\bibfnamefont {G.}~\bibnamefont {Dorda}},\
  and\ \bibinfo {author} {\bibfnamefont {M.}~\bibnamefont {Pepper}},\ }\href
  {https://doi.org/10.1103/physrevlett.45.494} {\bibfield  {journal} {\bibinfo
  {journal} {Physical Review Letters}\ }\textbf {\bibinfo {volume} {45}},\
  \bibinfo {pages} {494} (\bibinfo {year} {1980})}\BibitemShut {NoStop}%
\bibitem [{\citenamefont {Tsui}, \citenamefont {Stormer},\ and\ \citenamefont
  {Gossard}(1982)}]{Tsui1982}%
  \BibitemOpen
  \bibfield  {author} {\bibinfo {author} {\bibfnamefont {D.~C.}\ \bibnamefont
  {Tsui}}, \bibinfo {author} {\bibfnamefont {H.~L.}\ \bibnamefont {Stormer}},\
  and\ \bibinfo {author} {\bibfnamefont {A.~C.}\ \bibnamefont {Gossard}},\
  }\href {https://doi.org/10.1103/physrevlett.48.1559} {\bibfield  {journal}
  {\bibinfo  {journal} {Physical Review Letters}\ }\textbf {\bibinfo {volume}
  {48}},\ \bibinfo {pages} {1559} (\bibinfo {year} {1982})}\BibitemShut
  {NoStop}%
\bibitem [{\citenamefont {Laughlin}(1983)}]{Laughlin1983}%
  \BibitemOpen
  \bibfield  {author} {\bibinfo {author} {\bibfnamefont {R.~B.}\ \bibnamefont
  {Laughlin}},\ }\href {https://doi.org/10.1103/physrevlett.50.1395} {\bibfield
   {journal} {\bibinfo  {journal} {Physical Review Letters}\ }\textbf {\bibinfo
  {volume} {50}},\ \bibinfo {pages} {1395} (\bibinfo {year}
  {1983})}\BibitemShut {NoStop}%
\bibitem [{\citenamefont {Jain}(1989)}]{Jain1989}%
  \BibitemOpen
  \bibfield  {author} {\bibinfo {author} {\bibfnamefont {J.~K.}\ \bibnamefont
  {Jain}},\ }\href {https://doi.org/10.1103/physrevlett.63.199} {\bibfield
  {journal} {\bibinfo  {journal} {Physical Review Letters}\ }\textbf {\bibinfo
  {volume} {63}},\ \bibinfo {pages} {199} (\bibinfo {year} {1989})}\BibitemShut
  {NoStop}%
\bibitem [{\citenamefont {Halperin}, \citenamefont {Lee},\ and\ \citenamefont
  {Read}(1993)}]{Halperin1993}%
  \BibitemOpen
  \bibfield  {author} {\bibinfo {author} {\bibfnamefont {B.~I.}\ \bibnamefont
  {Halperin}}, \bibinfo {author} {\bibfnamefont {P.~A.}\ \bibnamefont {Lee}},\
  and\ \bibinfo {author} {\bibfnamefont {N.}~\bibnamefont {Read}},\ }\href
  {https://doi.org/10.1103/physrevb.47.7312} {\bibfield  {journal} {\bibinfo
  {journal} {Physical Review B}\ }\textbf {\bibinfo {volume} {47}},\ \bibinfo
  {pages} {7312} (\bibinfo {year} {1993})}\BibitemShut {NoStop}%
\bibitem [{\citenamefont {Stormer}, \citenamefont {Tsui},\ and\ \citenamefont
  {Gossard}(1999)}]{Stormer1999}%
  \BibitemOpen
  \bibfield  {author} {\bibinfo {author} {\bibfnamefont {H.~L.}\ \bibnamefont
  {Stormer}}, \bibinfo {author} {\bibfnamefont {D.~C.}\ \bibnamefont {Tsui}},\
  and\ \bibinfo {author} {\bibfnamefont {A.~C.}\ \bibnamefont {Gossard}},\
  }\href {https://doi.org/10.1103/revmodphys.71.s298} {\bibfield  {journal}
  {\bibinfo  {journal} {Reviews of Modern Physics}\ }\textbf {\bibinfo {volume}
  {71}},\ \bibinfo {pages} {S298} (\bibinfo {year} {1999})}\BibitemShut
  {NoStop}%
\bibitem [{\citenamefont {Konig}\ \emph {et~al.}(2007)\citenamefont {Konig},
  \citenamefont {Wiedmann}, \citenamefont {Bruune}, \citenamefont {Roth},
  \citenamefont {Buhmann}, \citenamefont {Molenkamp}, \citenamefont {Qi},\ and\
  \citenamefont {Zhang}}]{Koenig2007}%
  \BibitemOpen
  \bibfield  {author} {\bibinfo {author} {\bibfnamefont {M.}~\bibnamefont
  {Konig}}, \bibinfo {author} {\bibfnamefont {S.}~\bibnamefont {Wiedmann}},
  \bibinfo {author} {\bibfnamefont {C.}~\bibnamefont {Bruune}}, \bibinfo
  {author} {\bibfnamefont {A.}~\bibnamefont {Roth}}, \bibinfo {author}
  {\bibfnamefont {H.}~\bibnamefont {Buhmann}}, \bibinfo {author} {\bibfnamefont
  {L.~W.}\ \bibnamefont {Molenkamp}}, \bibinfo {author} {\bibfnamefont {X.-L.}\
  \bibnamefont {Qi}},\ and\ \bibinfo {author} {\bibfnamefont {S.-C.}\
  \bibnamefont {Zhang}},\ }\href {https://doi.org/10.1126/science.1148047}
  {\bibfield  {journal} {\bibinfo  {journal} {Science}\ }\textbf {\bibinfo
  {volume} {318}},\ \bibinfo {pages} {766} (\bibinfo {year}
  {2007})}\BibitemShut {NoStop}%
\bibitem [{\citenamefont {Bernevig}\ and\ \citenamefont
  {Zhang}(2006)}]{Bernevig2006}%
  \BibitemOpen
  \bibfield  {author} {\bibinfo {author} {\bibfnamefont {B.~A.}\ \bibnamefont
  {Bernevig}}\ and\ \bibinfo {author} {\bibfnamefont {S.-C.}\ \bibnamefont
  {Zhang}},\ }\href {https://doi.org/10.1103/physrevlett.96.106802} {\bibfield
  {journal} {\bibinfo  {journal} {Physical Review Letters}\ }\textbf {\bibinfo
  {volume} {96}},\ \bibinfo {pages} {106802} (\bibinfo {year}
  {2006})}\BibitemShut {NoStop}%
\bibitem [{\citenamefont {Bernevig}, \citenamefont {Hughes},\ and\
  \citenamefont {Zhang}(2006)}]{Bernevig2006a}%
  \BibitemOpen
  \bibfield  {author} {\bibinfo {author} {\bibfnamefont {B.~A.}\ \bibnamefont
  {Bernevig}}, \bibinfo {author} {\bibfnamefont {T.~L.}\ \bibnamefont
  {Hughes}},\ and\ \bibinfo {author} {\bibfnamefont {S.-C.}\ \bibnamefont
  {Zhang}},\ }\href {https://doi.org/10.1126/science.1133734} {\bibfield
  {journal} {\bibinfo  {journal} {Science}\ }\textbf {\bibinfo {volume}
  {314}},\ \bibinfo {pages} {1757} (\bibinfo {year} {2006})}\BibitemShut
  {NoStop}%
\bibitem [{\citenamefont {Konig}\ \emph {et~al.}(2008)\citenamefont {Konig},
  \citenamefont {Buhmann}, \citenamefont {Molenkamp}, \citenamefont {Hughes},
  \citenamefont {Liu}, \citenamefont {Qi},\ and\ \citenamefont
  {Zhang}}]{Koenig2008}%
  \BibitemOpen
  \bibfield  {author} {\bibinfo {author} {\bibfnamefont {M.}~\bibnamefont
  {Konig}}, \bibinfo {author} {\bibfnamefont {H.}~\bibnamefont {Buhmann}},
  \bibinfo {author} {\bibfnamefont {L.~W.}\ \bibnamefont {Molenkamp}}, \bibinfo
  {author} {\bibfnamefont {T.}~\bibnamefont {Hughes}}, \bibinfo {author}
  {\bibfnamefont {C.-X.}\ \bibnamefont {Liu}}, \bibinfo {author} {\bibfnamefont
  {X.-L.}\ \bibnamefont {Qi}},\ and\ \bibinfo {author} {\bibfnamefont {S.-C.}\
  \bibnamefont {Zhang}},\ }\href {https://doi.org/10.1143/jpsj.77.031007}
  {\bibfield  {journal} {\bibinfo  {journal} {Journal of the Physical Society
  of Japan}\ }\textbf {\bibinfo {volume} {77}},\ \bibinfo {pages} {031007}
  (\bibinfo {year} {2008})}\BibitemShut {NoStop}%
\bibitem [{\citenamefont {Chang}\ \emph {et~al.}(2013)\citenamefont {Chang},
  \citenamefont {Zhang}, \citenamefont {Feng}, \citenamefont {Shen},
  \citenamefont {Zhang}, \citenamefont {Guo}, \citenamefont {Li}, \citenamefont
  {Ou}, \citenamefont {Wei}, \citenamefont {Wang}, \citenamefont {Ji},
  \citenamefont {Feng}, \citenamefont {Ji}, \citenamefont {Chen}, \citenamefont
  {Jia}, \citenamefont {Dai}, \citenamefont {Fang}, \citenamefont {Zhang},
  \citenamefont {He}, \citenamefont {Wang}, \citenamefont {Lu}, \citenamefont
  {Ma},\ and\ \citenamefont {Xue}}]{Chang2013}%
  \BibitemOpen
  \bibfield  {author} {\bibinfo {author} {\bibfnamefont {C.-Z.}\ \bibnamefont
  {Chang}}, \bibinfo {author} {\bibfnamefont {J.}~\bibnamefont {Zhang}},
  \bibinfo {author} {\bibfnamefont {X.}~\bibnamefont {Feng}}, \bibinfo {author}
  {\bibfnamefont {J.}~\bibnamefont {Shen}}, \bibinfo {author} {\bibfnamefont
  {Z.}~\bibnamefont {Zhang}}, \bibinfo {author} {\bibfnamefont
  {M.}~\bibnamefont {Guo}}, \bibinfo {author} {\bibfnamefont {K.}~\bibnamefont
  {Li}}, \bibinfo {author} {\bibfnamefont {Y.}~\bibnamefont {Ou}}, \bibinfo
  {author} {\bibfnamefont {P.}~\bibnamefont {Wei}}, \bibinfo {author}
  {\bibfnamefont {L.-L.}\ \bibnamefont {Wang}}, \bibinfo {author}
  {\bibfnamefont {Z.-Q.}\ \bibnamefont {Ji}}, \bibinfo {author} {\bibfnamefont
  {Y.}~\bibnamefont {Feng}}, \bibinfo {author} {\bibfnamefont {S.}~\bibnamefont
  {Ji}}, \bibinfo {author} {\bibfnamefont {X.}~\bibnamefont {Chen}}, \bibinfo
  {author} {\bibfnamefont {J.}~\bibnamefont {Jia}}, \bibinfo {author}
  {\bibfnamefont {X.}~\bibnamefont {Dai}}, \bibinfo {author} {\bibfnamefont
  {Z.}~\bibnamefont {Fang}}, \bibinfo {author} {\bibfnamefont {S.-C.}\
  \bibnamefont {Zhang}}, \bibinfo {author} {\bibfnamefont {K.}~\bibnamefont
  {He}}, \bibinfo {author} {\bibfnamefont {Y.}~\bibnamefont {Wang}}, \bibinfo
  {author} {\bibfnamefont {L.}~\bibnamefont {Lu}}, \bibinfo {author}
  {\bibfnamefont {X.-C.}\ \bibnamefont {Ma}},\ and\ \bibinfo {author}
  {\bibfnamefont {Q.-K.}\ \bibnamefont {Xue}},\ }\href
  {https://doi.org/10.1126/science.1234414} {\bibfield  {journal} {\bibinfo
  {journal} {Science}\ }\textbf {\bibinfo {volume} {340}},\ \bibinfo {pages}
  {167} (\bibinfo {year} {2013})}\BibitemShut {NoStop}%
\bibitem [{\citenamefont {Liu}, \citenamefont {Zhang},\ and\ \citenamefont
  {Qi}(2016)}]{Liu2016}%
  \BibitemOpen
  \bibfield  {author} {\bibinfo {author} {\bibfnamefont {C.-X.}\ \bibnamefont
  {Liu}}, \bibinfo {author} {\bibfnamefont {S.-C.}\ \bibnamefont {Zhang}},\
  and\ \bibinfo {author} {\bibfnamefont {X.-L.}\ \bibnamefont {Qi}},\ }\href
  {https://doi.org/10.1146/annurev-conmatphys-031115-011417} {\bibfield
  {journal} {\bibinfo  {journal} {Annual Review of Condensed Matter Physics}\
  }\textbf {\bibinfo {volume} {7}},\ \bibinfo {pages} {301} (\bibinfo {year}
  {2016})}\BibitemShut {NoStop}%
\bibitem [{\citenamefont {Tang}, \citenamefont {Mei},\ and\ \citenamefont
  {Wen}(2011)}]{Tang2011}%
  \BibitemOpen
  \bibfield  {author} {\bibinfo {author} {\bibfnamefont {E.}~\bibnamefont
  {Tang}}, \bibinfo {author} {\bibfnamefont {J.-W.}\ \bibnamefont {Mei}},\ and\
  \bibinfo {author} {\bibfnamefont {X.-G.}\ \bibnamefont {Wen}},\ }\href
  {https://doi.org/10.1103/physrevlett.106.236802} {\bibfield  {journal}
  {\bibinfo  {journal} {Physical Review Letters}\ }\textbf {\bibinfo {volume}
  {106}},\ \bibinfo {pages} {236802} (\bibinfo {year} {2011})}\BibitemShut
  {NoStop}%
\bibitem [{\citenamefont {Rod}, \citenamefont {Schmidt},\ and\ \citenamefont
  {Rachel}(2015)}]{Rod2015}%
  \BibitemOpen
  \bibfield  {author} {\bibinfo {author} {\bibfnamefont {A.}~\bibnamefont
  {Rod}}, \bibinfo {author} {\bibfnamefont {T.~L.}\ \bibnamefont {Schmidt}},\
  and\ \bibinfo {author} {\bibfnamefont {S.}~\bibnamefont {Rachel}},\ }\href
  {https://doi.org/10.1103/physrevb.91.245112} {\bibfield  {journal} {\bibinfo
  {journal} {Physical Review B}\ }\textbf {\bibinfo {volume} {91}},\ \bibinfo
  {pages} {245112} (\bibinfo {year} {2015})}\BibitemShut {NoStop}%
\bibitem [{\citenamefont {Neupert}\ \emph
  {et~al.}(2011{\natexlab{a}})\citenamefont {Neupert}, \citenamefont {Santos},
  \citenamefont {Chamon},\ and\ \citenamefont {Mudry}}]{Neupert2011}%
  \BibitemOpen
  \bibfield  {author} {\bibinfo {author} {\bibfnamefont {T.}~\bibnamefont
  {Neupert}}, \bibinfo {author} {\bibfnamefont {L.}~\bibnamefont {Santos}},
  \bibinfo {author} {\bibfnamefont {C.}~\bibnamefont {Chamon}},\ and\ \bibinfo
  {author} {\bibfnamefont {C.}~\bibnamefont {Mudry}},\ }\href
  {https://doi.org/10.1103/physrevlett.106.236804} {\bibfield  {journal}
  {\bibinfo  {journal} {Physical Review Letters}\ }\textbf {\bibinfo {volume}
  {106}},\ \bibinfo {pages} {236804} (\bibinfo {year}
  {2011}{\natexlab{a}})}\BibitemShut {NoStop}%
\bibitem [{\citenamefont {Neupert}\ \emph
  {et~al.}(2011{\natexlab{b}})\citenamefont {Neupert}, \citenamefont {Santos},
  \citenamefont {Ryu}, \citenamefont {Chamon},\ and\ \citenamefont
  {Mudry}}]{Neupert2011a}%
  \BibitemOpen
  \bibfield  {author} {\bibinfo {author} {\bibfnamefont {T.}~\bibnamefont
  {Neupert}}, \bibinfo {author} {\bibfnamefont {L.}~\bibnamefont {Santos}},
  \bibinfo {author} {\bibfnamefont {S.}~\bibnamefont {Ryu}}, \bibinfo {author}
  {\bibfnamefont {C.}~\bibnamefont {Chamon}},\ and\ \bibinfo {author}
  {\bibfnamefont {C.}~\bibnamefont {Mudry}},\ }\href
  {https://doi.org/10.1103/physrevb.84.165107} {\bibfield  {journal} {\bibinfo
  {journal} {Physical Review B}\ }\textbf {\bibinfo {volume} {84}},\ \bibinfo
  {pages} {165107} (\bibinfo {year} {2011}{\natexlab{b}})}\BibitemShut
  {NoStop}%
\bibitem [{\citenamefont {Neupert}\ \emph
  {et~al.}(2012{\natexlab{a}})\citenamefont {Neupert}, \citenamefont {Santos},
  \citenamefont {Ryu}, \citenamefont {Chamon},\ and\ \citenamefont
  {Mudry}}]{Neupert2012}%
  \BibitemOpen
  \bibfield  {author} {\bibinfo {author} {\bibfnamefont {T.}~\bibnamefont
  {Neupert}}, \bibinfo {author} {\bibfnamefont {L.}~\bibnamefont {Santos}},
  \bibinfo {author} {\bibfnamefont {S.}~\bibnamefont {Ryu}}, \bibinfo {author}
  {\bibfnamefont {C.}~\bibnamefont {Chamon}},\ and\ \bibinfo {author}
  {\bibfnamefont {C.}~\bibnamefont {Mudry}},\ }\href
  {https://doi.org/10.1103/physrevlett.108.046806} {\bibfield  {journal}
  {\bibinfo  {journal} {Physical Review Letters}\ }\textbf {\bibinfo {volume}
  {108}},\ \bibinfo {pages} {046806} (\bibinfo {year}
  {2012}{\natexlab{a}})}\BibitemShut {NoStop}%
\bibitem [{\citenamefont {Neupert}\ \emph
  {et~al.}(2012{\natexlab{b}})\citenamefont {Neupert}, \citenamefont {Santos},
  \citenamefont {Chamon},\ and\ \citenamefont {Mudry}}]{Neupert2012a}%
  \BibitemOpen
  \bibfield  {author} {\bibinfo {author} {\bibfnamefont {T.}~\bibnamefont
  {Neupert}}, \bibinfo {author} {\bibfnamefont {L.}~\bibnamefont {Santos}},
  \bibinfo {author} {\bibfnamefont {C.}~\bibnamefont {Chamon}},\ and\ \bibinfo
  {author} {\bibfnamefont {C.}~\bibnamefont {Mudry}},\ }\href
  {https://doi.org/10.1103/physrevb.86.165133} {\bibfield  {journal} {\bibinfo
  {journal} {Physical Review B}\ }\textbf {\bibinfo {volume} {86}},\ \bibinfo
  {pages} {165133} (\bibinfo {year} {2012}{\natexlab{b}})}\BibitemShut
  {NoStop}%
\bibitem [{\citenamefont {Grushin}\ \emph {et~al.}(2012)\citenamefont
  {Grushin}, \citenamefont {Neupert}, \citenamefont {Chamon},\ and\
  \citenamefont {Mudry}}]{Grushin2012}%
  \BibitemOpen
  \bibfield  {author} {\bibinfo {author} {\bibfnamefont {A.~G.}\ \bibnamefont
  {Grushin}}, \bibinfo {author} {\bibfnamefont {T.}~\bibnamefont {Neupert}},
  \bibinfo {author} {\bibfnamefont {C.}~\bibnamefont {Chamon}},\ and\ \bibinfo
  {author} {\bibfnamefont {C.}~\bibnamefont {Mudry}},\ }\href
  {https://doi.org/10.1103/physrevb.86.205125} {\bibfield  {journal} {\bibinfo
  {journal} {Physical Review B}\ }\textbf {\bibinfo {volume} {86}},\ \bibinfo
  {pages} {205125} (\bibinfo {year} {2012})}\BibitemShut {NoStop}%
\bibitem [{\citenamefont {Xie}\ \emph {et~al.}(2021)\citenamefont {Xie},
  \citenamefont {Pierce}, \citenamefont {Park}, \citenamefont {Parker},
  \citenamefont {Khalaf}, \citenamefont {Ledwith}, \citenamefont {Cao},
  \citenamefont {Lee}, \citenamefont {Chen}, \citenamefont {Forrester},
  \citenamefont {Watanabe}, \citenamefont {Taniguchi}, \citenamefont
  {Vishwanath}, \citenamefont {Jarillo-Herrero},\ and\ \citenamefont
  {Yacoby}}]{Xie2021}%
  \BibitemOpen
  \bibfield  {author} {\bibinfo {author} {\bibfnamefont {Y.}~\bibnamefont
  {Xie}}, \bibinfo {author} {\bibfnamefont {A.~T.}\ \bibnamefont {Pierce}},
  \bibinfo {author} {\bibfnamefont {J.~M.}\ \bibnamefont {Park}}, \bibinfo
  {author} {\bibfnamefont {D.~E.}\ \bibnamefont {Parker}}, \bibinfo {author}
  {\bibfnamefont {E.}~\bibnamefont {Khalaf}}, \bibinfo {author} {\bibfnamefont
  {P.}~\bibnamefont {Ledwith}}, \bibinfo {author} {\bibfnamefont
  {Y.}~\bibnamefont {Cao}}, \bibinfo {author} {\bibfnamefont {S.~H.}\
  \bibnamefont {Lee}}, \bibinfo {author} {\bibfnamefont {S.}~\bibnamefont
  {Chen}}, \bibinfo {author} {\bibfnamefont {P.~R.}\ \bibnamefont {Forrester}},
  \bibinfo {author} {\bibfnamefont {K.}~\bibnamefont {Watanabe}}, \bibinfo
  {author} {\bibfnamefont {T.}~\bibnamefont {Taniguchi}}, \bibinfo {author}
  {\bibfnamefont {A.}~\bibnamefont {Vishwanath}}, \bibinfo {author}
  {\bibfnamefont {P.}~\bibnamefont {Jarillo-Herrero}},\ and\ \bibinfo {author}
  {\bibfnamefont {A.}~\bibnamefont {Yacoby}},\ }\href
  {https://doi.org/10.1038/s41586-021-04002-3} {\bibfield  {journal} {\bibinfo
  {journal} {Nature}\ }\textbf {\bibinfo {volume} {600}},\ \bibinfo {pages}
  {439} (\bibinfo {year} {2021})}\BibitemShut {NoStop}%
\bibitem [{\citenamefont {Cheng}\ \emph {et~al.}(2019)\citenamefont {Cheng},
  \citenamefont {Pan}, \citenamefont {Che}, \citenamefont {Wang}, \citenamefont
  {Wu}, \citenamefont {Watanabe}, \citenamefont {Taniguchi}, \citenamefont
  {Ge}, \citenamefont {Lake}, \citenamefont {Smirnov}, \citenamefont {Lau},\
  and\ \citenamefont {Bockrath}}]{Cheng2019}%
  \BibitemOpen
  \bibfield  {author} {\bibinfo {author} {\bibfnamefont {B.}~\bibnamefont
  {Cheng}}, \bibinfo {author} {\bibfnamefont {C.}~\bibnamefont {Pan}}, \bibinfo
  {author} {\bibfnamefont {S.}~\bibnamefont {Che}}, \bibinfo {author}
  {\bibfnamefont {P.}~\bibnamefont {Wang}}, \bibinfo {author} {\bibfnamefont
  {Y.}~\bibnamefont {Wu}}, \bibinfo {author} {\bibfnamefont {K.}~\bibnamefont
  {Watanabe}}, \bibinfo {author} {\bibfnamefont {T.}~\bibnamefont {Taniguchi}},
  \bibinfo {author} {\bibfnamefont {S.}~\bibnamefont {Ge}}, \bibinfo {author}
  {\bibfnamefont {R.}~\bibnamefont {Lake}}, \bibinfo {author} {\bibfnamefont
  {D.}~\bibnamefont {Smirnov}}, \bibinfo {author} {\bibfnamefont {C.~N.}\
  \bibnamefont {Lau}},\ and\ \bibinfo {author} {\bibfnamefont {M.}~\bibnamefont
  {Bockrath}},\ }\href {https://doi.org/10.1021/acs.nanolett.9b00811}
  {\bibfield  {journal} {\bibinfo  {journal} {Nano Letters}\ }\textbf {\bibinfo
  {volume} {19}},\ \bibinfo {pages} {4321} (\bibinfo {year}
  {2019})}\BibitemShut {NoStop}%
\bibitem [{\citenamefont {Yin}\ \emph {et~al.}(2020)\citenamefont {Yin},
  \citenamefont {Ma}, \citenamefont {Cochran}, \citenamefont {Xu},
  \citenamefont {Zhang}, \citenamefont {Tien}, \citenamefont {Shumiya},
  \citenamefont {Cheng}, \citenamefont {Jiang}, \citenamefont {Lian},
  \citenamefont {Song}, \citenamefont {Chang}, \citenamefont {Belopolski},
  \citenamefont {Multer}, \citenamefont {Litskevich}, \citenamefont {Cheng},
  \citenamefont {Yang}, \citenamefont {Swidler}, \citenamefont {Zhou},
  \citenamefont {Lin}, \citenamefont {Neupert}, \citenamefont {Wang},
  \citenamefont {Yao}, \citenamefont {Chang}, \citenamefont {Jia},\ and\
  \citenamefont {Hasan}}]{Yin2020}%
  \BibitemOpen
  \bibfield  {author} {\bibinfo {author} {\bibfnamefont {J.-X.}\ \bibnamefont
  {Yin}}, \bibinfo {author} {\bibfnamefont {W.}~\bibnamefont {Ma}}, \bibinfo
  {author} {\bibfnamefont {T.~A.}\ \bibnamefont {Cochran}}, \bibinfo {author}
  {\bibfnamefont {X.}~\bibnamefont {Xu}}, \bibinfo {author} {\bibfnamefont
  {S.~S.}\ \bibnamefont {Zhang}}, \bibinfo {author} {\bibfnamefont {H.-J.}\
  \bibnamefont {Tien}}, \bibinfo {author} {\bibfnamefont {N.}~\bibnamefont
  {Shumiya}}, \bibinfo {author} {\bibfnamefont {G.}~\bibnamefont {Cheng}},
  \bibinfo {author} {\bibfnamefont {K.}~\bibnamefont {Jiang}}, \bibinfo
  {author} {\bibfnamefont {B.}~\bibnamefont {Lian}}, \bibinfo {author}
  {\bibfnamefont {Z.}~\bibnamefont {Song}}, \bibinfo {author} {\bibfnamefont
  {G.}~\bibnamefont {Chang}}, \bibinfo {author} {\bibfnamefont
  {I.}~\bibnamefont {Belopolski}}, \bibinfo {author} {\bibfnamefont
  {D.}~\bibnamefont {Multer}}, \bibinfo {author} {\bibfnamefont
  {M.}~\bibnamefont {Litskevich}}, \bibinfo {author} {\bibfnamefont {Z.-J.}\
  \bibnamefont {Cheng}}, \bibinfo {author} {\bibfnamefont {X.~P.}\ \bibnamefont
  {Yang}}, \bibinfo {author} {\bibfnamefont {B.}~\bibnamefont {Swidler}},
  \bibinfo {author} {\bibfnamefont {H.}~\bibnamefont {Zhou}}, \bibinfo {author}
  {\bibfnamefont {H.}~\bibnamefont {Lin}}, \bibinfo {author} {\bibfnamefont
  {T.}~\bibnamefont {Neupert}}, \bibinfo {author} {\bibfnamefont
  {Z.}~\bibnamefont {Wang}}, \bibinfo {author} {\bibfnamefont {N.}~\bibnamefont
  {Yao}}, \bibinfo {author} {\bibfnamefont {T.-R.}\ \bibnamefont {Chang}},
  \bibinfo {author} {\bibfnamefont {S.}~\bibnamefont {Jia}},\ and\ \bibinfo
  {author} {\bibfnamefont {M.~Z.}\ \bibnamefont {Hasan}},\ }\href
  {https://doi.org/10.1038/s41586-020-2482-7} {\bibfield  {journal} {\bibinfo
  {journal} {Nature}\ }\textbf {\bibinfo {volume} {583}},\ \bibinfo {pages}
  {533} (\bibinfo {year} {2020})}\BibitemShut {NoStop}%
\bibitem [{\citenamefont {Sims}(2022)}]{Sims2022}%
  \BibitemOpen
  \bibfield  {author} {\bibinfo {author} {\bibfnamefont {C.}~\bibnamefont
  {Sims}},\ }\href {https://doi.org/10.3390/condmat7020040} {\bibfield
  {journal} {\bibinfo  {journal} {Condensed Matter}\ }\textbf {\bibinfo
  {volume} {7}},\ \bibinfo {pages} {40} (\bibinfo {year} {2022})}\BibitemShut
  {NoStop}%
\bibitem [{\citenamefont {Kane}\ and\ \citenamefont {Mele}(2005)}]{Kane2005}%
  \BibitemOpen
  \bibfield  {author} {\bibinfo {author} {\bibfnamefont {C.~L.}\ \bibnamefont
  {Kane}}\ and\ \bibinfo {author} {\bibfnamefont {E.~J.}\ \bibnamefont
  {Mele}},\ }\href {https://doi.org/10.1103/physrevlett.95.226801} {\bibfield
  {journal} {\bibinfo  {journal} {Physical Review Letters}\ }\textbf {\bibinfo
  {volume} {95}},\ \bibinfo {pages} {226801} (\bibinfo {year}
  {2005})}\BibitemShut {NoStop}%
\bibitem [{\citenamefont {Liu}, \citenamefont {Jiang},\ and\ \citenamefont
  {Yao}(2011)}]{Liu2011}%
  \BibitemOpen
  \bibfield  {author} {\bibinfo {author} {\bibfnamefont {C.-C.}\ \bibnamefont
  {Liu}}, \bibinfo {author} {\bibfnamefont {H.}~\bibnamefont {Jiang}},\ and\
  \bibinfo {author} {\bibfnamefont {Y.}~\bibnamefont {Yao}},\ }\href
  {https://doi.org/10.1103/physrevb.84.195430} {\bibfield  {journal} {\bibinfo
  {journal} {Physical Review B}\ }\textbf {\bibinfo {volume} {84}},\ \bibinfo
  {pages} {195430} (\bibinfo {year} {2011})}\BibitemShut {NoStop}%
\bibitem [{\citenamefont {Ezawa}(2013{\natexlab{a}})}]{Ezawa2013}%
  \BibitemOpen
  \bibfield  {author} {\bibinfo {author} {\bibfnamefont {M.}~\bibnamefont
  {Ezawa}},\ }\href {https://doi.org/10.1038/srep03435} {\bibfield  {journal}
  {\bibinfo  {journal} {Scientific Reports}\ }\textbf {\bibinfo {volume} {3}}
  (\bibinfo {year} {2013}{\natexlab{a}}),\ 10.1038/srep03435}\BibitemShut
  {NoStop}%
\bibitem [{\citenamefont {Ezawa}(2013{\natexlab{b}})}]{Ezawa2013a}%
  \BibitemOpen
  \bibfield  {author} {\bibinfo {author} {\bibfnamefont {M.}~\bibnamefont
  {Ezawa}},\ }\href {https://doi.org/10.1103/physrevb.87.155415} {\bibfield
  {journal} {\bibinfo  {journal} {Physical Review B}\ }\textbf {\bibinfo
  {volume} {87}},\ \bibinfo {pages} {155415} (\bibinfo {year}
  {2013}{\natexlab{b}})}\BibitemShut {NoStop}%
\bibitem [{\citenamefont {Kim}\ \emph {et~al.}(2016)\citenamefont {Kim},
  \citenamefont {Ochoa}, \citenamefont {Zarzuela},\ and\ \citenamefont
  {Tserkovnyak}}]{Kim2016}%
  \BibitemOpen
  \bibfield  {author} {\bibinfo {author} {\bibfnamefont {S.~K.}\ \bibnamefont
  {Kim}}, \bibinfo {author} {\bibfnamefont {H.}~\bibnamefont {Ochoa}}, \bibinfo
  {author} {\bibfnamefont {R.}~\bibnamefont {Zarzuela}},\ and\ \bibinfo
  {author} {\bibfnamefont {Y.}~\bibnamefont {Tserkovnyak}},\ }\href
  {https://doi.org/10.1103/physrevlett.117.227201} {\bibfield  {journal}
  {\bibinfo  {journal} {Physical Review Letters}\ }\textbf {\bibinfo {volume}
  {117}},\ \bibinfo {pages} {227201} (\bibinfo {year} {2016})}\BibitemShut
  {NoStop}%
\bibitem [{\citenamefont {Regnault}\ and\ \citenamefont
  {Bernevig}(2011)}]{Regnault2011}%
  \BibitemOpen
  \bibfield  {author} {\bibinfo {author} {\bibfnamefont {N.}~\bibnamefont
  {Regnault}}\ and\ \bibinfo {author} {\bibfnamefont {B.~A.}\ \bibnamefont
  {Bernevig}},\ }\href {https://doi.org/10.1103/physrevx.1.021014} {\bibfield
  {journal} {\bibinfo  {journal} {Physical Review X}\ }\textbf {\bibinfo
  {volume} {1}},\ \bibinfo {pages} {021014} (\bibinfo {year}
  {2011})}\BibitemShut {NoStop}%
\end{thebibliography}%
%aipnum4-2.bst 2019-01-14 (MD) hand-edited version of apsrev4-1.bst
%Control: key (0)
%Control: author (8) initials jnrlst
%Control: editor formatted (1) identically to author
%Control: production of article title (-1) disabled
%Control: page (0) single
%Control: year (1) truncated
%Control: production of eprint (0) enabled
%

%%%%%%%%%%%%%%%%%%%%%%%%%%%%%%%%%%%%%%%%%%
\end{document}